%% file: paper.tex
\documentclass[twocolumn,showpacs,aps,prl,superscriptaddress]{revtex4}

\usepackage{graphicx}
\usepackage{dcolumn}
\usepackage{amsmath}
\usepackage{epsfig}

\input babarsym

\newcommand{\BABARPubYear}    {11}
\newcommand{\BABARPubNumber}  {005}
\newcommand{\SLACPubNumber} {14470}

\def\figurebox#1#2#3{%
    \def\arg{#3}%
    \ifx\arg\empty
    {\hfill\vbox{\hsize#2\hrule\hbox to #2{\vrule\hfill\vbox to #1{\hsize#2\vfill}\vrule}\hrule}\hfill}%
    \else
    {\hfill\epsfbox{#3}\hfill}%
    \fi}

\begin{document}

\begin{flushleft}
\babar-PUB-\BABARPubYear/\BABARPubNumber \\
SLAC-PUB-\SLACPubNumber \\
\end{flushleft}

\title{
  {\large \bf
    Study of di-pion bottomonium transitions and 
    search for the $h_b(1P)$ state 
  }
}

\input authors_mar2011.tex

\date{\today}

\begin{abstract}
We study inclusive di-pion decays using a sample of $108\times 10^6$  
$\Upsilon(3S)$ events recorded with the \babar\ detector. 
We search for the decay mode $\Upsilon(3S)\to \pi^+\pi^- h_b(1P)$ and find no evidence for the bottomonium spin-singlet state $h_b(1P)$ in the invariant mass distribution recoiling against the $\pi^+\pi^-$ system.
Assuming the $h_b(1P)$ mass to be $9.900\,\gevcc$, 
we measure the upper limit on the branching fraction ${\cal B}[\Upsilon(3S)\to \pi^+ \pi^- h_b(1P)] < 1.2 \times 10^{-4}$, at 90\% confidence level.
We also investigate the $\chi_{bJ}(2P)\to \pi^+\pi^-\chi_{bJ}(1P)$, $\Upsilon(3S)\to \pi^+\pi^-\Upsilon(2S)$, and  $\Upsilon(2S)\to \pi^+\pi^-\Upsilon(1S)$ di-pion transitions and 
present an improved measurement of the branching fraction of the $\Upsilon(3S)\to \pi^+\pi^-\Upsilon(2S)$ decay 
and of the $\Upsilon(3S)-\Upsilon(2S)$ mass difference. 

\end{abstract}


\maketitle

Studies of $b\bar{b}$ ({\it bottomonium}) and $c\bar{c}$ 
({\it charmonium}) bound states provide insight about inter-quark forces.
The measurement of the hyperfine mass splitting between 
triplet and singlet states in quarkonium systems 
discriminates between various models and tests lattice QCD and potential nonrelativistic QCD %pNRQCD 
calculations~\cite{ref:models}.
Observation of the $P$-wave singlet ground state of charmonium, $h_c(1P)$, was recently    
confirmed  
and its mass precisely measured, yielding the hyperfine splitting for the charmonium $1P$ states 
$\Delta M_{hf}(1P)_{c\bar{c}}\equiv \langle M({}^3P_J)\rangle_{c\bar{c}} - M({}^1P_1)_{c\bar{c}}=+0.08\pm 0.18{\rm (stat.)}\pm 0.12{\rm (syst.)}\,\mevcc$~\cite{ref:hc}, where $\langle M({}^3P_J)\rangle$ is the 
spin-weighted average mass of the $J=0,1,2$ ground states.  
The hyperfine splitting $\Delta M_{hf}(1P)_{b\bar{b}}\equiv  \langle M({}^3P_J)\rangle_{b\bar{b}} - M({}^1P_1)_{b\bar{b}}$ for  
bottomonium states is expected to be no more than a few \mevcc~\cite{ref:godfrey}.
The ${}^3P_J$ $b\bar{b}$ ground states are well-established, and their    
spin-weighted mass average is 
$\langle M({}^3P_J)_{b\bar{b}}\rangle = [M(\chi_{b0}(1P))+3M(\chi_{b1}(1P))+5M(\chi_{b2}(1P))]/9=9.89987\pm 0.00027\,\gevcc$~\cite{ref:pdg}.
The $h_b(1P)$, hereafter referred to as the $h_b$, is expected to decay predominantly to $ggg$ ($57\%$ branching fraction), $\gamma\eta_b$ ($41\%$), and $gg\gamma$ ($2\%$), 
and its width 
is predicted to be of order $0.1\,\mev$~\cite{ref:GR02}.

We report, herein, a search for the $h_b$ through the % 
hadronic transition $\Upsilon(3S) \to \pi^+ \pi^- h_b(1P)$. 
The CLEO Collaboration searched for the $h_b$ in the reactions 
$\Upsilon(3S)\to\pi^0 h_b$ and $\Upsilon(3S)\to\pi^+\pi^- h_b$, setting  
upper limits at 90\% confidence level (CL)
for the branching fractions %
${\cal B}[\Upsilon(3S)\to\pi^0 h_b]<2.7\times 10^{-3}$ and 
${\cal B}_{\Upsilon}\equiv{\cal B}[\Upsilon(3S)\to\pi^+\pi^- h_b]<1.8\times 10^{-3}$, 
assuming the $h_b$ mass $m(h_b)$ to be $9.900\,\gevcc$~\cite{ref:cleohb}.
The \babar\ Collaboration recently reported evidence for the $h_b$ in $\Upsilon(3S)\to \pi^0 h_b$, 
$h_b\to \eta_b\gamma$ decays~\cite{ref:babarpiz}. 
Preliminary results of a search for the $h_b$ in the reaction $e^+ e^- \to \pi^+ \pi^- h_b$, reporting the observation of the $h_b$ meson, have been announced by the Belle Collaboration~\cite{ref:bellehb}. 
Theoretical predictions for 
${\cal B}_{\Upsilon}$ span  
one order of magnitude. References~\cite{ref:standardmultipoleTuan,ref:Kuang,ref:kuangyan} 
predict a branching fraction between $2.2\times 10^{-4}$ and $8.0\times 10^{-4}$, 
while Ref.~\cite{ref:Voloshin}
predicts a rate of $10^{-4}$ or smaller.

The data sample used in this study was collected with the \babar\ 
detector~\cite{ref:babar} 
at the PEP-II asymmetric-energy $e^+e^-$ storage rings at the SLAC National 
Accelerator Laboratory.
It consists of $25.6\,\invfb$ of integrated luminosity collected at a 
$e^+e^-$ center-of-mass (c.m.) energy of $10.355\,\gev$, corresponding to the 
mass of the $\Upsilon(3S)$ resonance. 
The number of recorded $\Upsilon(3S)$ events is $108\times 10^6$.
An additional sample of $2.5\,\invfb$ recorded at the $\Upsilon(3S)$ energy (``10\%'' sample)
and a $2.6\,\invfb$ sample collected $30\,\mev$ below the $\Upsilon(3S)$ resonance (``off-peak'' sample)  
are used for background and calibration studies. 

The momenta of charged particles are reconstructed
using a combination of a five-layer double-sided silicon-strip 
detector and a 40-layer drift chamber, both operating in the 1.5-T magnetic field
of a superconducting solenoid.
Photons are detected using a CsI(Tl) electromagnetic calorimeter, which is also inside the magnet coil.
Charged hadron identification 
is achieved through measurements of particle energy loss in 
the tracking system and the Cherenkov angle obtained from a detector of internally reflected Cherenkov light.

The $\pi^+\pi^-$ pairs are selected from oppositely-charged tracks that    
originate from the \epem\ interaction region in hadronic events, 
hence excluding tracks arising
from a photon conversion or the decay of a long-lived particle. We 
search for an 
$h_b$ signal  
using a fit to the spectrum of the mass $m_R$ recoiling against the 
$\pi^+\pi^-$ system, defined by: 
\begin{equation}
  m_R^2= (M_{\Upsilon(3S)}-E^*_{\pi\pi})^2-|{\mathbf P}^{*}_{\pi\pi}|^{2},  
\end{equation} 
where $E^*_{\pi\pi}$ and ${\mathbf P}^*_{\pi\pi}$ are, respectively, the measured $\pi\pi$ energy and momentum in the c.m. frame.

The $h_b$ signal is expected to appear as a peak in the $m_R$ distribution on top of a smooth non-peaking 
background from continuum events ($e^+e^- \to q\bar q$ with $q=u,d,s,c$) and
bottomonium decays.
Several other processes produce peaks in the recoil mass spectrum 
close to the signal region.
Hadronic transitions 
$\Upsilon(3S)\to\pi^+\pi^-\Upsilon(2S)$ 
(hereafter denoted $\Upsilon^{3\to 2}$) produce 
a peak centered at the $\Upsilon(2S)$ mass $m[\Upsilon(2S)]=10.02326 \pm 0.00031\,\gevcc$~\cite{ref:pdg}.
The cascade process 
$\Upsilon(3S)\to\Upsilon(2S)X$, 
$\Upsilon(2S)\to\pi^+\pi^-\Upsilon(1S)$ ($\Upsilon^{2\to 1}$)
results in a peak centered at $9.791\,\gevcc$. %
The peak is offset from the $\Upsilon(1S)$ mass by approximately the 
$\Upsilon(3S)$ to $\Upsilon(2S)$ mass difference.
Doppler shift and broadening further affect the position and width of this peak.
When the $\Upsilon(2S)$ parent 
in $\Upsilon^{2\to 1}$ decays 
is produced through the $\Upsilon^{3\to 2}$ channel, 
a pion from the $\Upsilon(3S)$ decay can be combined with an 
oppositely-charged  
track from the $\Upsilon(2S)$ decay to produce a broad distribution 
centered around $9.9\,\gevcc$.
The $\Upsilon(2S)$ is also produced through the 
initial-state radiation (ISR) process $e^+e^-\to \gamma_{ISR} \Upsilon(2S)$ 
($\Upsilon^{2\to 1}_{ISR}$). % 
Of the nine possible $\Upsilon(3S)\to\chi_{bJ'}(2P)\gamma$, $\chi_{bJ'}(2P)\to\pi^+\pi^-\chi_{bJ}(1P)$ decay chains ($\chi_b^{J',J}$)~\cite{ref:chibKuang}, 
only those for $J'=J=\{1,2\}$ have been reported~\cite{ref:pdg,ref:CawlfieldCLEO}; these should generate two narrowly separated peaks near $9.993\,\gevcc$, 
while the contributions with $J'\neq J$ or with $J=0$ are expected to be negligible.

Selection criteria are chosen by optimizing
the ratio $S/\sqrt{B}$ between the expected $h_b$  signal yield ($S$) and
the background ($B$). The signal sample for the optimization
is provided by a detailed Monte Carlo (MC) simulation based on \textsc{Geant4}~\cite{ref:geant}, 
EvtGen~\cite{ref:evtgen}, and \textsc{Jetset}~\cite{ref:jetset},  
while the background sample is obtained from the 10\% sample, 
which is not used for the extraction of the signal.  
The natural width of the $h_b$ meson, which is predicted to be negligible 
in comparison with the experimental resolution in $m_R$ ($0.009\,\gevcc$ r.m.s.), is 
set to zero in the simulation. 

Since decays of the $h_{b}$ via three gluons or to $\eta_b\gamma$  
are expected to exhibit a high track multiplicity,  
we require an event to have between  
$6$ and $16$ charged tracks, where the upper restriction reduces contributions due to random combinations of tracks. 
We further require the ratio of the second to zeroth 
Fox-Wolfram moment~\cite{ref:fox} calculated using all  
charged tracks and unmatched neutral showers in the event 
to be less than 0.55.  
 The total event energy in the laboratory frame must lie between $6$ and $18$\,\gev, where the lower restriction reduces QED background.
%

%%%%
Events must contain two oppositely-charged tracks, each of which is 
 identified as a pion.
The pion identification efficiency depends on momentum and polar angle, and 
is typically about $98\%$.  
This requirement provides a rejection factor of order 50
against electrons. 
The vertex of each reconstructed pion pair must lie within 
$0.41\,$cm and be less than $4\sigma_L$ from the nominal interaction point  
in the transverse plane,
where $\sigma_L$ is the  
uncertainty evaluated on a candidate-by-candidate basis for the transverse flight length $L$.
We demand the $\chi^2$ probability for the vertex fit to be greater than $0.001$.

%%%%

%%%
The phase-space distribution of \KS\ decays extends up to 
$m_R$ values of approximately 
$9.86\,\gevcc$ and then rapidly decreases.
To further suppress the background due to \KS\ decays, we reject pairs 
of pions if their vertex is displaced from the nominal interaction point  
by more than $0.05\,$cm and $2\sigma_L$ in the transverse plane and 
if they satisfy 
$\cos\alpha>0.95$, where $\alpha$ is the angle between the direction of the di-pion candidate momentum 
and its flight direction in the transverse plane.
Candidates  
removed from the nominal sample that satisfy 
all other selection criteria constitute a \KS-enriched control sample.

The selected data sample consists of approximately $137\times 10^6$ $\pi^+\pi^-$ candidates  in the range $9.750<m_R<10.040\,$\gevcc, corresponding to an average of $2.4$ selected di-pion candidates per event.
The fit validation studies described below account for the effect of
candidate multiplicity. 
We evaluate the di-pion reconstruction efficiency with MC events, by  
matching the reconstructed $\pi^+\pi^-$ pairs to the simulated pairs  
emitted in the bottomonium transition under study on an event-by-event basis.  
The $h_b$ signal efficiency is $\epsilon=41.8\%$ for 
$m(h_b)=9.900\,\gevcc$, with a $[+6,-3]\%$ variation of $\epsilon$ over the $m(h_b)$ range $[9.880,9.920]\,\gevcc$.  
A lower reconstruction efficiency of $25.0\%$ ($16.7\%$) is found for the softer $\pi^+\pi^-$ pairs 
produced in $\chi_b^{J',J}$ ($\Upsilon^{3\to 2}$) transitions.
For the $\Upsilon^{2\to 1}$ transition, an efficiency of $47.2\%$ is obtained by averaging over the contributions from $X=\gamma\gamma$, $\pi^0\pi^0$, and $\pi^+\pi^-$.

 We perform a $\chi^2$ fit to the $m_R$ spectrum 
in the range $9.750<m_R<10.040\,$\gevcc with a model comprising  
eight components: 
non-peaking background, 
$\Upsilon^{3\to 2}$,  
$\Upsilon^{2\to 1}$, $\Upsilon^{2\to 1}_{ISR}$,  
$\chi_b^{2,2}$, $\chi_b^{1,1}$,  
$\KS\to\pi^+\pi^-$,
and the $h_b$ signal.  
The $m_R$ distributions of the signal and background are parameterized  
using probability density functions (PDFs).
We define a two-sided Crystal Ball (TCB) function, which is a Gaussian 
for $-\alpha_L < (x-x_0)/\sigma < \alpha_R$, and transitions to the power-law tail function $f(x)$~\cite{ref:CrystalBall}: 
\begin{equation}
f(x)=e^{-\frac{1}{2}\alpha_{i}^2}{\displaystyle\left({\frac{n_{i}}{\alpha_{i}}}\right)^{n_{i}}\left/\left(\frac{|x - x_0|}{\sigma}+\frac{n_{i}}{\alpha_{i}}-\alpha_{i}\right)^{n_{i}}\right. },
\label{eq:twosidedcb}
\end{equation}
 where $x_0$ and $\sigma$ are the mean and width of the Gaussian, and 
the subscript $i=L$ ($i=R$) applies to values $x<x_0$ ($x>x_0$).
We model the signal component with a symmetric ($\alpha_L=\alpha_R$, $n_L=n_R$) TCB shape. 

The $\Upsilon^{3\to 2}$ and  
$\Upsilon^{2\to 1}$ peaks 
are described by 
sums of an asymmetric TCB shape and an asymmetric Gaussian. Contributions to 
$\Upsilon^{2\to 1}$ from $X=\{\pi^+\pi^-,\pi^0\pi^0,\gamma\gamma\}$ are modeled  
separately because of the different Doppler broadening. 
Their relative fractions and relative peak positions are fixed according to the 
world-average values~\cite{ref:pdg} and the MC-simulated 
$m_R$ spectrum, respectively.
For each peak, the ratios of the widths of the TCB and Gaussian functions 
are fixed to the values found from fitting the corresponding MC spectrum.
The PDF of the peaking background from ISR $\Upsilon(2S)$ production 
is parameterized as a symmetric TCB function  
whose parameters are determined from simulated events.
The yield of $\Upsilon^{2\to 1}_{ISR}$ events in the  $\Upsilon(3S)$ sample, 
$[6.6 \pm 1.0{\rm(stat.)}]\times 10^4$, is determined using the off-peak data.
A symmetric TCB function is used as the PDF for both the $\chi_{b}^{1,1}$ and 
$\chi_{b}^{2,2}$ contributions. 
The peak positions of the $\chi_{b}^{J',J}$ components  
relative to the $\Upsilon^{3\to 2}$ peak are % 
fixed according to the MC simulation.
The parameters for the width and tail of the TCB function
are common to  
both $\chi_{b}^{J',J}$ peaks.

The \KS\ background is modeled using empirical phase space 
functions derived from the MC. 
Knowledge about the \KS\ transverse momentum  
distribution is 
obtained from fits to the $\pi^+\pi^-$ invariant mass spectrum for the 
\KS-enriched sample, 
and is used to  
correct discrepancies between the data and the MC simulation. 
The \KS\ background yield, $(348 \pm 10)\times 10^3$, 
is obtained from an extrapolation of a fit to the $m_R$ spectrum of the \KS-enriched  
sample, using a scale factor of $2.5$ determined from MC simulation. 
The non-peaking background PDF 
is parameterized by a sixth-order 
polynomial.

The signal (peaking background) PDF 
excludes  
random combinations of tracks  
that do not originate from the signal (background) bottomonium transition.  
Such misreconstructed combinations are included in the non-peaking term.

To improve fit stability, the fit is performed in two stages: a preliminary fit to fix background parameters 
followed by a final fit. 
The peaking background PDF parameters and yields are determined from the  
preliminary, $\chi^2$-based fit in which the signal component is 
excluded from the model. 
The free parameters in the fit are: the yields of the continuum background 
and the peaking background components $\Upsilon^{3\to 2}$, 
$\Upsilon^{2\to 1}$, and $\chi_b^{J',J}$; 
the continuum background PDF parameters; 
the overall $m_R$ scale of the \KS\ contribution; the peak positions 
of the  
$\Upsilon^{3\to 2}$ and  
$\Upsilon^{2\to 1}$ components;  
the overall widths of the PDFs for the $\Upsilon^{3\to 2}$, 
$\Upsilon^{2\to 1}$, and $\chi_b^{J',J}$ components. 
The $\chi^2$ per degree of freedom after the preliminary fit is 
$364/272 \approx 1.3$, 
where the largest contributions arise from a few isolated 
bins near $9.79$ and $10.02\,\gevcc$.
As the measurement is dominated by systematic uncertainties, we 
evaluate the $\chisq$ distribution on simulated pseudoexperiments accounting for the 
dominant sources of systematic uncertainties, and we observe values of  
$\chisq$ greater than $364$ in more than $7\%$ of the trials.
In the final fit, all peaking-background parameters except the yields 
are fixed to the values extracted from the preliminary fit.

The final fit is performed as a scan over the values 
of the $h_b$ peak position, with $39$ steps in $1\,\mevcc$ intervals in the range 
$(9.880,9.920)\,\gevcc$.
At each step, a $\chi^2$ fit is performed for the signal and background 
yields and the continuum background parameters.
The fit procedure is validated with 
simulated experiments,  
and systematic uncertainties are evaluated for each point of the scan.

\begin{figure}[b]
\begin{center}
\includegraphics[height=2.2in]{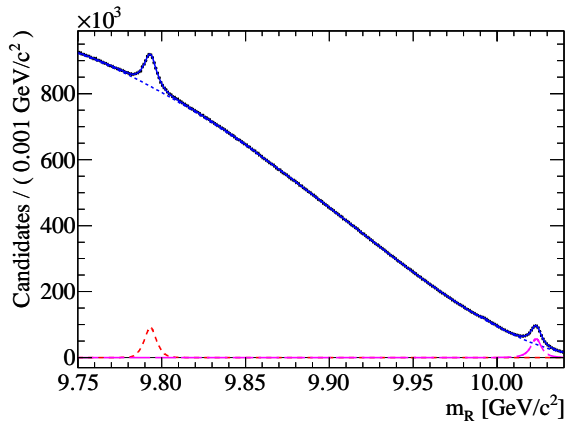}
\end{center}
\caption{\label{fig:spectrum} 
The $m_R$ spectrum: data (points) with 
  the fitted model (solid line) superimposed. The short-dashed line is the 
  contribution from the continuum component.
Also shown are the $\Upsilon^{2\to 1}$ (dashed curve around $9.79\,\gevcc$)
and $\Upsilon^{3\to 2}$ (long-dashed curve around $10.02\,\gevcc$) components.
The $h_b$ signal component is excluded from the superimposed model. 
} 
\end{figure}

\begin{figure}[bt]
\begin{center}
\includegraphics[height=2.2in]{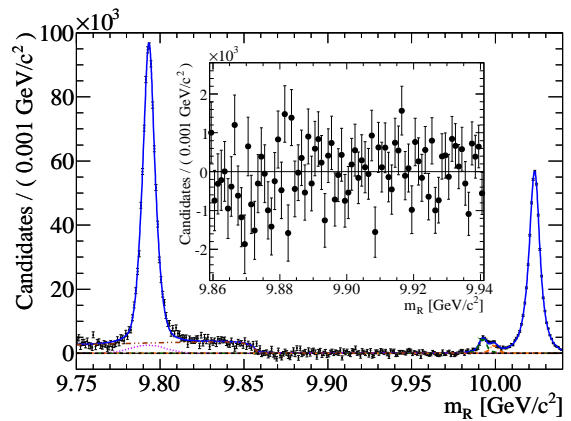}
\end{center}
\caption{\label{fig:subspectrum} 
The $m_R$ spectrum after subtraction of the continuum 
  background component.  
The curves represent the  fitted model (solid), the 
  $\Upsilon^{2\to 1}_{ISR}$ (dotted),
  \KS\ (double-dot-dashed), 
  $\chi_b^{1,1}$ (dashed), 
  and $\chi_b^{2,2}$ (dot-dashed) components. 
  Inset: expanded view in the $h_b$  
region  
  after subtraction of continuum and peaking backgrounds. 
}
\end{figure}

\begin{figure}[bt]
\begin{center}
\includegraphics[height=2.2in]{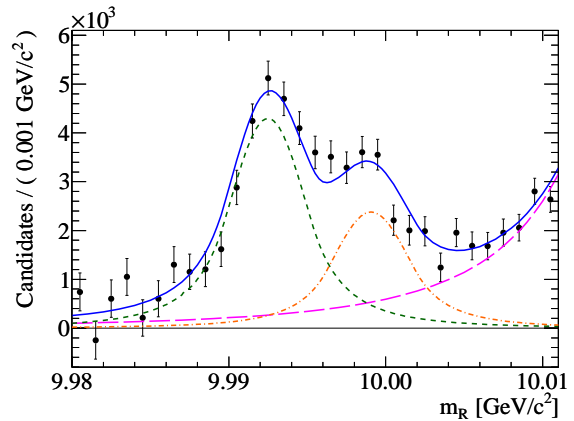}
\end{center}
\caption{\label{fig:zoomchi} 
The $m_R$ spectrum in the $\chi_b^{J',J}$ region 
  after subtraction of continuum and \KS\ background components: 
  points represent data, while the curves represent the fitted model (solid), the $\chi_b^{1,1}$ 
  (dashed), $\chi_b^{2,2}$ 
  (dot-dashed), and $\Upsilon^{3\to 2}$ 
  (long-dashed) components.
}
\end{figure}

Figure~\ref{fig:spectrum} shows the $m_R$ spectrum and the fit result. 
The non-peaking background component dominates, 
with only the prominent $\Upsilon^{3\to 2}$ and $\Upsilon^{2\to 1}$ 
peaks clearly seen above this background.  
When comparing the fitted mass of the $\Upsilon^{3\to 2}$ 
peak with MC simulation, we observe a $+0.44\pm 0.02{\rm (stat.)}\,\mevcc$ displacement in data, which 
corresponds to a difference of $331.50 \pm 0.02{\rm (stat.)} \pm 0.13{\rm (syst.)}\,\mevcc$ 
between the $\Upsilon(3S)$ and $\Upsilon(2S)$ masses. 
The systematic uncertainty is dominated by uncertainties in the lineshape and in the track momentum measurement. 
Details of the latter may be found in Ref.~\cite{ref:taumass}. Other sources 
of uncertainty have been investigated and found to be of minor significance. These 
include the fit bias, the c.m. boost relative to the laboratory, the 
background model, and the PDF parameters. 
The position of the $\Upsilon^{2\to 1}$ peak is shifted by $+1.23\pm
0.02{\rm (stat.)}\,\mevcc$  in data with respect to 
simulation and corresponds to a difference of $561.7 \pm 0.0{\rm (stat.)}  \pm 1.2{\rm (syst.)}\,\mevcc$ 
between the $\Upsilon(2S)$ and $\Upsilon(1S)$ masses, where the dominant sources of systematic
uncertainty are the $\Upsilon(2S)$ momentum in the c.m. frame and the
lineshape model. 
Figure~\ref{fig:subspectrum} shows  the distribution of $m_R$ 
after subtraction of the non-peaking background.   
An expanded view of the $\chi_b^{J',J}$ region is presented 
in Fig.~\ref{fig:zoomchi}. 

The inset of Fig.~\ref{fig:subspectrum} shows an expanded view of the $h_b$ signal region.
The significance of a signal is evaluated at each point of the scan using the ratio, $N/\sigma_N$,  of 
the signal yield $N$ over its uncertainty $\sigma_N$. 
The largest enhancement over background is a $2.2$ standard deviation ($\sigma$) excess (statistical only) 
at $m(h_b)\approx 9.916\,\gevcc$. Therefore, we do not obtain evidence for an $h_b$ signal.
The fitted $h_b$ signal yield for $m(h_b)=9.900\,\gevcc$ 
 is $[-1.1 \pm 2.4{\rm (stat.)}]\times 10^3$ events. % 
Results for the $\Upsilon^{3\to 2}$, $\Upsilon^{2\to 1}$, 
$\chi_{b}^{1\to 1}$, and $\chi_b^{2\to 2}$ product branching fractions 
are presented in Table~\ref{tab:bbres}.

\begin{table*}[t]
\caption{\label{tab:bbres}
  Summary of results for the signal yields, reconstruction efficiency
  $\epsilon$, uncertainties on the branching fraction due to fit bias,
    systematic uncertainties on yields and efficiencies, the product branching fraction 
and the branching fraction for the di-pion transition. 
}
\begin{center}
\begin{tabular}{l|c@{\hspace{4pt}}c@{\hspace{8pt}}c@{\hspace{8pt}}c@{\hspace{8pt}}c}\hline\noalign{\vskip 1pt}
 & $h_b$ [$m(h_b)=9.900\,\gevcc$]& $\chi_b^{1\to 1}$ & $\chi_b^{2\to 2}$ & $\Upsilon^{3\to 2}$ &  $\Upsilon^{2\to 1}$  \\ \noalign{\vskip 1pt}\hline
Yield		& $-1106 \pm 2432$ & $31418\pm 1851$ & $17385\pm 1456$ & $543839\pm 2928 $& $906059\pm 7407 $\\
$\epsilon$ (\%) &  41.8     & 25.0 &  25.0 & 16.7 & 47.2\\ 
Fit bias $(10^{-3})$ &  $-0.06$    & $-0.09$ &$-0.04$ & $+0.2$ & $+0.8$ \\
Yield error $(10^{-3})$ &  $0.01$   & $0.06$ & $0.06$& $0.6$ & $0.4$ \\ %  
$\epsilon$ error $(10^{-3})$ & $0.00$   & $0.05$ & $0.03$ & $1.3$ & $0.8$ \\ 
$\prod {\cal B}$ $(10^{-3})$	& ...  &  $1.16\pm 0.07\pm 0.12$ & $0.64\pm 0.05\pm 0.08$ & ... & $17.8\pm 0.2 \pm 1.1$ \\ 
${\cal B}$ $(10^{-3})$	&  $-0.02 \pm 0.05 \pm 0.06$  & $9.2 \pm 0.6 \pm 0.9$ & $4.9 \pm 0.4 \pm 0.6$ & $30.0\pm 0.2 \pm 1.4$ & ... \\ 
 \hline \hline
\end{tabular}
\end{center}
\end{table*}
 
In the following, reported quantities refer to $m(h_b)=9.900\,\gevcc$. 
The ranges spanned by varying $m(h_b)$ in the interval $[9.880,9.920]\,\gevcc$ are given in parentheses.
The following systematic uncertainties are associated with the signal yields. 
We observe a $10\%$ discrepancy between the $m_R$ resolution values in data and MC for the 
$\Upsilon^{3\to 2}$ component, and translate this into an uncertainty of $0.1\times 10^3$ ($0.0\times 10^3$ to $0.4\times 10^3$) events on the $h_b$ signal yield.   
A systematic uncertainty of $0.4\times 10^3$ ($0.3\times 10^3$ to $0.5\times 10^3$)  
events is estimated by 
varying the PDF parameters fixed in the fit by $\pm$1 $\sigma$, 
varying the overall width of the $h_b$ PDF by $10\%$, 
 setting the yield of the ISR $\Upsilon(2S)$ component to  
 $\pm$1 $\sigma$ of the nominal value, and    
varying the \KS\ component normalization and parameters 
within their uncertainties. 
Uncertainties related to the continuum background model 
 amount to $0.2\times 10^3$ ($0.0\times 10^3$ to $0.7\times 10^3$) % 
events.
The additive systematic uncertainties on the yields of the $\Upsilon^{3\to 2}$, $\chi_{b}^{1\to 1}$ 
and $\chi_b^{2\to 2}$ components also account for the 
modeling of the $\Upsilon^{3\to 2}$ tails and for the assumption that the 
contributions of the $\Upsilon(3S)\to X \chi_{bJ'}(2P)$, $\chi_{bJ'}(2P)\to \pi^+\pi^- \chi_{bJ}$ decay chains with $J'\neq J$ or $J=0$ are
negligible~\cite{ref:chibKuang,ref:CawlfieldCLEO}.

The fit bias on the extracted yields, due to the choice of the fit model, 
is estimated with pseudoexperiments based on fully simulated Monte Carlo 
samples. 
  We estimate a fit bias on the $h_b$ signal yield of $-2.8\times 10^3$ ($-3.0\times 10^3$ to $+0.4\times 10^3$)  
  events. 
Fit biases for the other di-pion transitions %, 
   are listed in Table~\ref{tab:bbres}.  
We do not correct the signal yields but rather assign the bias 
as a systematic uncertainty.

The following systematic uncertainties are associated with the reconstruction efficiency $\epsilon$. 
The uncertainty due to the track-reconstruction efficiency is $3\%$.
To assess the impact of data-MC differences  
on the $\pi^+\pi^-$ candidate selection efficiencies, 
we compare the relative variations of the $\Upsilon^{3\to 2}$
   yield in data and MC when excluding selection requirements one at a time,
and assign the full observed discrepancy to the systematic uncertainty. 
A total uncertainty of $2.3\%$ in $\epsilon$ is obtained,  %
including the statistical uncertainty ($0.6\%$) in the $\Upsilon^{3\to 2}$ 
yield.
The uncertainty in the number of $\Upsilon(3S)$ events is $1.1\%$. %
The above multiplicative systematic uncertainties affect 
the product branching fractions of all di-pion transitions studied in this analysis. %
Differences in the selection efficiencies resulting from different 
angular distributions of the $h_b$ decay products and 
different $h_b$ hadronization models in the MC simulation 
contribute a $5\%$ uncertainty. 
Model uncertainties in the simulation of the di-pion kinematics, bottomonium hadronization, and $\Upsilon(2S)$ production channel (where applicable) in the   
$\Upsilon^{2\to 1}$, $\Upsilon^{3\to 2}$~\cite{ref:CroninCLEO}, $\chi_b^{1\to 1}$, and $\chi_b^{2\to 2}$ decay 
chains result in systematic uncertainties on the efficiency of 
$1.3\%$, $0.5\%$, $0.6\%$, and $0.6\%$, respectively.

\begin{figure}[bt]
\begin{center}
\includegraphics[height=2.2in]{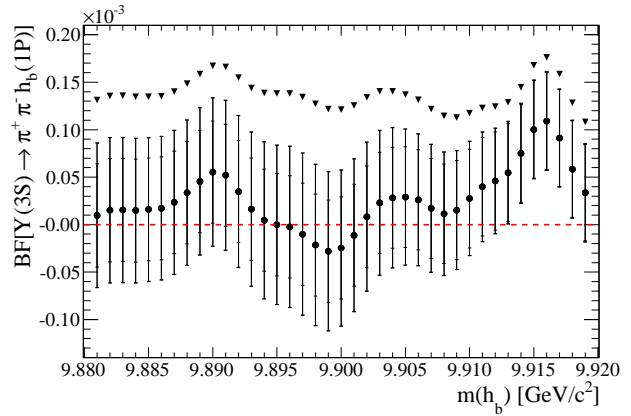}
\end{center}
\caption{\label{fig:scan}
Results for the $\Upsilon(3S)\to\pi^+\pi^-h_b$ branching fraction (points with statistical and systematic uncertainties) as a function of the assumed $h_b$ mass. The triangles indicate the upper limits at 90\% CL. 
}
\end{figure}

Product branching fractions
are calculated by  
dividing the fitted yield by the efficiency  
and the % 
 number 
of $\Upsilon(3S)$ events,  %  
and are summarized in Table~\ref{tab:bbres}. 
For $m(h_b)=9.900\,\gevcc$ we find the branching fraction ${\cal B}_{\Upsilon} \equiv {\cal B}[\Upsilon(3S)\to\pi^+\pi^-h_b] = (-0.2 \pm 0.5 \pm 0.6) \times 10^{-4}$,  
 where the first uncertainty is statistical and the second systematic, 
and set an upper limit (UL) ${\cal B}_{\Upsilon} <1.2 \times 10^{-4}$ at 90\% CL. 
The UL is calculated assuming a Gaussian sampling distribution 
$f({\cal B}_{\Upsilon})$ for ${\cal B}_{\Upsilon}$, which accounts for statistical and systematic uncertainties, 
and determining the value of UL for 
which  
$\int_{0}^{UL}f({\cal B}_{\Upsilon})d{\cal B}_{\Upsilon} = 0.9 \times \int_{0}^{\infty}f({\cal B}_{\Upsilon})d{\cal B}_{\Upsilon}$. 
Figure~\ref{fig:scan} reports the branching fractions ${\cal B}_{\Upsilon}$ and the 
90\% CL ULs as a function of the assumed $h_b$ mass.
% %
The branching fractions of the $\chi_{b1}(2P)\to\pi^+\pi^-\chi_{b1}$ and 
$\chi_{b2}(2P)\to\pi^+\pi^-\chi_{b2}$ transitions, given in Table~\ref{tab:bbres}, 
are derived by correcting for the branching fractions of the  
$\Upsilon(3S)\to \gamma \chi_{b1}(2P)$ and $\Upsilon(3S)\to \gamma \chi_{b2}(2P)$ decays~\cite{ref:pdg, ref:ArtusoCLEO}, respectively. 
The extracted values are in reasonable agreement with those found in the study by the CLEO Collaboration~\cite{ref:pdg,ref:CawlfieldCLEO}, where the two transitions could not be separated experimentally. 

In summary, 
we present an inclusive analysis of the $\pi^+\pi^-$ recoil 
mass spectrum in $\Upsilon(3S)$ decays.  
We measure the branching fraction 
\begin{align*}
\lefteqn{{\cal B}[\Upsilon(3S)\to\pi^+\pi^-\Upsilon(2S)]}\\
&&=(3.00\pm 0.02{\rm (stat.)}\pm 0.14{\rm (syst.)})\%. 
\end{align*}
This value is in reasonable agreement with, and more precise than, the current world average $(2.45\pm 0.23)\%$~\cite{ref:pdg}.

The measured $\Upsilon(3S)$-$\Upsilon(2S)$ mass difference is $331.50 \pm 0.02 ({\rm stat.}) \pm 0.13({\rm syst.})\,\mevcc$. 

We extract the product branching fractions 
\begin{align*}
{\cal B}[\Upsilon(3S)\to X\chi_{b1}(2P)]\times {\cal B}[\chi_{b1}(2P)\to\pi^+\pi^-\chi_{b1}]\\
= (1.16\pm 0.07\pm 0.12)\times 10^{-3}, 
\end{align*}
\begin{align*}
{\cal B}[\Upsilon(3S)\to X\chi_{b2}(2P)]\times {\cal B}[\chi_{b2}(2P)\to\pi^+\pi^-\chi_{b2}]\\
= (0.64\pm 0.05\pm 0.08)\times 10^{-3}, \quad {\rm and} 
\end{align*}
\begin{align*}
{\cal B}[\Upsilon(3S)\to X\Upsilon(2S)]\times {\cal B}[\Upsilon(2S)\to\pi^+\pi^-\Upsilon]\\
= (1.78\pm 0.02\pm 0.11)\%. 
\end{align*}
A search for the $h_b$ state, the ${}^1P_1$ state of bottomonium, in $\Upsilon(3S) \to \pi^+\pi^- h_b$ decays 
does not provide evidence for this decay mode, and assuming 
the $h_b$ mass to be $9.900\,\gevcc$, we set a $90\%$ CL upper limit 
${\cal B}_{\Upsilon}<1.2 \times 10^{-4}$.
We exclude, at 
90\% CL, values of ${\cal B}_{\Upsilon}$ above $1.8\times 10^{-4}$ 
for a wide range of assumed $h_b$ mass values. 
%%%
These results disfavor the calculations 
of Refs.~\cite{ref:standardmultipoleTuan,ref:Kuang,ref:kuangyan}. 
Similarly, a recent measurement of the 
$\Upsilon(1^3D_J)\to\Upsilon(1S)\pi^+\pi^-$ branching fraction~\cite{ref:bab1d}
disfavors the calculations of Ref.~\cite{ref:Kuang,ref:kuangyan}.
% %
The predictions of Ref.~\cite{ref:Voloshin} are 
at least one order of magnitude smaller and are not contradicted by our result.

\input acknow.tex

\end{document}

%% file: authors_mar2011.tex
%% author list as of 01-Mar-2011 (389 authors)
%
\author{J.~P.~Lees}
\author{V.~Poireau}
\author{V.~Tisserand}
\affiliation{Laboratoire d'Annecy-le-Vieux de Physique des Particules (LAPP), Universit\'e de Savoie, CNRS/IN2P3,  F-74941 Annecy-Le-Vieux, France}
\author{J.~Garra~Tico}
\author{E.~Grauges}
\affiliation{Universitat de Barcelona, Facultat de Fisica, Departament ECM, E-08028 Barcelona, Spain }
\author{M.~Martinelli$^{ab}$}
\author{D.~A.~Milanes$^{a}$}
\author{A.~Palano$^{ab}$ }
\author{M.~Pappagallo$^{ab}$ }
\affiliation{INFN Sezione di Bari$^{a}$; Dipartimento di Fisica, Universit\`a di Bari$^{b}$, I-70126 Bari, Italy }
\author{G.~Eigen}
\author{B.~Stugu}
\author{L.~Sun}
\affiliation{University of Bergen, Institute of Physics, N-5007 Bergen, Norway }
% ...
\collaboration{The \babar\ Collaboration}
\noaffiliation

%% file: acknow.tex
We are grateful for the excellent luminosity and machine conditions
provided by our \pep2\ colleagues, 
and for the substantial dedicated effort from
the computing organizations that support \babar. 
The collaborating institutions wish to thank 
SLAC for its support and kind hospitality. 
This work is supported by
DOE
and NSF (USA),
NSERC (Canada),
CEA and
CNRS-IN2P3
(France),
BMBF and DFG
(Germany),
INFN (Italy),
FOM (The Netherlands),
NFR (Norway),
MES (Russia),
MEC (Spain), and
STFC (United Kingdom). 
Individuals have received support from the
Marie Curie EIF (European Union) and
the A.~P.~Sloan Foundation.